\documentstyle[aps,prc,epsf,twocolumn]{revtex}
\begin{document}
\title{Analysis of the instantaneous Bethe-Salpeter
       equation for \(q\bar{q}\)-bound-states}
\author{J.Resag, C.R.M\"unz, B.C.Metsch, H.R.Petry}
\address{Institut f\"ur Theoretische Kernphysik,\\
         Universit\"at Bonn, Nussallee 14-16, 53115 Bonn, FRG}
\date{\today}
\maketitle

\begin{abstract}
We investigate the structure of the instantaneous
Bethe-Salpeter equation for \(q\bar{q}\)-bound states
in the general case of unequal quark masses
and develop a numerical scheme for the calculation of
mass spectra and Bethe-Salpeter amplitudes.
In order to appreciate the merits of the
various competing models beyond the reproduction of the mass spectra
we present explicit formulas to calculate electroweak decays.
The results for an explicit quark model will be compared
to experimental data in a subsequent paper \cite{Mue}.
\end{abstract} \pacs{}

\narrowtext

\section{Introduction} \label{I}
Despite many efforts the bound state problem of QCD is still far
from being well understood. One of the main tasks is to investigate
the relevance of quarks as dynamical degrees of freedom in hadronic bound
states. Since a relativistic treatment for the quarks in deeply bound states is
essential, the Bethe-Salpeter(BS)-equation \cite{BS,GL}
provides a suitable starting point.
Unlike in QED one cannot use perturbation theory
to obtain useful approximations for the interaction kernel in QCD.
Therefore our knowledge of the interaction between quarks is still
quite fragmentary and various phenomenological alternatives have
to be tested. In the present work we will restrict ourselves
to \(q\bar{q}\)-states. The use of general \(q\bar{q}\)-interaction
kernels depending
on the relative time variable leads to serious
conceptional and practical problems \cite{Nak,Nk2}. Therefore it
is very useful at this point
to make the simplifying assumption that the BS-kernel can be approximated
by an effective interaction that is instantaneous in the rest frame
of the bound state. The BS-equation then reduces to the (full) Salpeter
equation \cite{Sa} that has been investigated for \(q\bar{q}\)-states
e.g. by Llewellyn Smith \cite{LS2}, Le Yaouanc and coworkers \cite{Ya}
and recently by Laga\"e \cite{La}.

In the present paper we analyze the properties
of the Salpeter equation for the general case of unequal quark masses
in Sec.\ref{II}.
In Sec.\ref{III} we present a flexible numerical
treatment of this equation based on the variational principle
of ref.\cite{La}. We use the block
structure of the Salpeter amplitude to derive an expansion
in terms of a complete set of basis functions
which leads to a matrix equation analogous
to the RPA (random phase approximation) equation well known in
nuclear theory \cite{RS}.
Our numerical method can be applied to a wide class of
phenomenological interaction kernels.
In Sec.\ref{IV} we show how to reconstruct the full BS-amplitudes
from the Salpeter amplitudes and present the calculation of some decay
observables within the Mandelstam formalism including the decay
\(\pi^0\rightarrow 2\gamma\). Concluding remarks given
are in Sec.\ref{V}.

The method obtained in this paper will be applied to an explicit
quark model for light mesons in a subsequent paper \cite{Mue}.

\section{General properties of the Salpeter equation} \label{II}
\subsection{Formulating the Salpeter equation} \label{IIA}
\begin{figure}[ht]
  \centering
  \leavevmode
  \epsfxsize=0.38\textwidth
  \epsffile{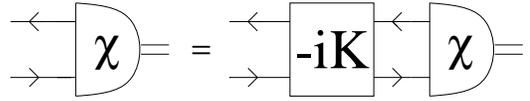}
\vspace{0.5cm}
\caption{Graphical representation of the BS equation}
\label{fig1}
\end{figure}

The BS-amplitude \(\chi\) for a fermion-antifermion bound state
\( \left| P \right\rangle \) is defined by
\begin{equation}
\left[ \chi_P(x_1,x_2) \right]_{\alpha\beta}
 = \left\langle\, 0 \,
\left|\, T\,\Psi_{\alpha}(x_1)\,\bar{\Psi}_{\beta}(x_2) \,
\right|\, P \,\right\rangle \label{1}
\end{equation}
where \(P\) is the four-momentum of the bound state,
\(T\) denotes the time ordering for the fermion operators
\(\Psi,\,\bar{\Psi}\) and \(\alpha,\,\beta\) stand for
spinor, flavor and color indices. Due to translational
invariance the Fourier transformation can be written as
\begin{equation}
\chi_P(x_1,x_2) = e^{-iPX}\,\int\frac{d^4p}{(2\pi)^4}\,
e^{-ipx}\,\chi_P(p) \label{2}
\end{equation}
where
\(x_1=X+\eta_2x ,\; x_2=X-\eta_1x \) with the conjugate momenta
\(  p_1=\eta_1P+p ,\; p_2=\eta_2P-p\).
Here \(\eta_1,\,\eta_2\) are two arbitrary
real numbers satisfying \(\eta_1+\eta_2=1\). The BS-equation for
\(\chi_P(p)\) then reads
\begin{eqnarray}
\lefteqn{\chi_P(p) = } \label{4}\\
&=& S^F_1(p_1)\,
\int \frac{d^4 p'}{(2\pi)^4}\,
[-i\,K(P,p,p')\,\chi_P(p')]\,
S^F_2(-p_2) \nonumber
\end{eqnarray}
also shown in graphical form in Fig.\ref{fig1}.
The interaction kernel \(K(P,p,p')\) generally acts
on \(\chi_P(p')\) as
\begin{equation}
[K(P,p,p')\,\chi_P(p')]_{\alpha\beta} = \sum_{\alpha'\,\beta'}\,
K(P,p,p')_{\alpha\alpha',\beta\beta'}\chi_P(p')_{\alpha'\,\beta'}
\end{equation}
For an interaction that is instantaneous in the rest frame
of the bound state with momentum \(P=(M,\vec{0})\)
the BS-kernel can be written as
\begin{equation}
K(P,p,p')\,\Big|_{P=(M,\vec{0})} = V(\vec{p},\vec{p}\,') \label{6}
\end{equation}
which can also be formulated in a covariant way as
\begin{equation}
K(P,p,p') = V(p_{\perp},\,p_{\perp}') \label{7}
\end{equation}
where
\(
p_{\perp} = p-(Pp/P^2)\,P
\)
is perpendicular to \(P\). In practical calculations
one has to justify this ansatz a posteriori by investigating its consequences
in the framework of explicit models.

Furthermore we will approximate the full quark propagators \(S^F(p)\) by
bare propagators
\(
S^F_i(p_i) \approx i\,\left( \gamma p_i - m_i + i\epsilon \right)^{-1}
\)
where \(m_1\) and \(m_2\) are interpreted as effective masses
for the quark and the antiquark. This approximation has been
criticized \cite{Bo} because free propagators might be incompatible
with a confining kernel. On the other hand one can argue that
this choice naturally leads to nonrelativistic potential models
that have been applied successfully to heavy quarkonia
(a recent model calculation is presented in \cite{Bey}, for an
extensive review see \cite{LSG}).
We thus feel that free propagators should be a reasonable effective
parameterization at least for heavy quarks. It is still an open
question whether free propagators can also be applied to light quarks,
and one has to investigate this problem within explicit models.

With an instantaneous BS-kernel and bare propagators with effective
quark masses one can perform the \(p^0\) integrals in the BS-equation
in the rest frame of the bound state with mass \(M\)
(see e.g. \cite{IZ}) and thus arrives at the (full) Salpeter equation
\begin{eqnarray}
\Phi(\vec{p}) &=&
\int \!\!\frac{d^3p'}{(2\pi)^3}\,
\frac{\Lambda^-_1(\vec{p})\,\gamma^0\,
[(V(\vec{p},\vec{p}\,')\,\Phi(\vec{p}\,')]
\,\gamma^0\,\Lambda^+_2(-\vec{p})}
{M+\omega_1+\omega_2}
 \nonumber \\
 &-&
\int \!\!\frac{d^3p'}{(2\pi)^3}\,
\frac{\Lambda^+_1(\vec{p})\,\gamma^0\,
[(V(\vec{p},\vec{p}\,')\,\Phi(\vec{p}\,')]
\,\gamma^0\,\Lambda^-_2(-\vec{p})}
{M-\omega_1-\omega_2}
 \label{9}
\end{eqnarray}
with \(\omega_i=\sqrt{\vec{p}\,^2+m_i^2}\) and the projection operators
\(
\Lambda^{\pm}_i(\vec{p}) = (\omega_i \pm H_i(\vec{p}))/(2\omega_i)
\) on positive and negative energies.
Here
\(H_i(\vec{p})=\gamma^0(\vec{\gamma}\vec{p}+m_i)\) is
the standard Dirac hamiltonian.
We also have introduced the Salpeter amplitude \(\Phi\) by
\begin{equation}
\Phi(\vec{p}) = \int dp^0\,\chi_P(p^0,\vec{p})\,\Big|_{P=(M,\vec{0})}
\label{12}
\end{equation}

For weakly bound states with \(|\vec{p}|/m_i \ll 1\) and
\(M \approx m_1+m_2\) one has
\begin{equation}
\frac{1}{M+\omega_1+\omega_2} \ll \frac{1}{M-\omega_1-\omega_2}
\label{13} \end{equation}
so that the first term in eq.(\ref{9}) can be dropped.
This leads to the so called reduced Salpeter equation,
which has been used
in various studies of relativistic bound states (see e.g. the work
of Gara et al. \cite{Ga} and references therein).
In the case of light quarks, however,
the use of the reduced Salpeter equation is dubious, especially for
deeply bound
states like the pion. Quark models for light quarks should therefore be based
on the full Salpeter equation eq.(\ref{9}).

Let \(\alpha,\,\beta\) in eq.(\ref{1}) refer to Dirac indices in the standard
Dirac representation of ref.\cite{IZ}.
Then \(\Phi\) is a 4\(\times\)4-matrix in spinor space
that can be written in block matrix form as
\begin{equation}
\Phi =
\left( \begin{array}{*{2}{c}}
\Phi^{+-} & \Phi^{++} \\
\Phi^{--} & \Phi^{-+}
\end{array} \right)
\label{14}
\end{equation}
where each component is a 2\(\times\)2-matrix.
Applying \(\Lambda^{\pm}_1(\vec{p})\) from the left hand side and
\(\Lambda^{\pm}_2(-\vec{p})\) from the right hand side to
the Salpeter equation leads to
\begin{eqnarray}
\Lambda^+_1(\vec{p})\,\Phi(\vec{p}) \,\Lambda^+_2(-\vec{p}) &=& 0
\nonumber \\
\Lambda^-_1(\vec{p})\,\Phi(\vec{p}) \,\Lambda^-_2(-\vec{p}) &=& 0
\label{15}
\end{eqnarray}
These relations allow us to express \(\Phi^{+-},\,\Phi^{-+}\) in terms of
\(\Phi^{++},\,\Phi^{--}\) as
\begin{eqnarray}
\Phi^{+-} &=& +c_1 \Phi^{++} s - c_2 s \Phi^{--} \nonumber \\
\Phi^{-+} &=& -c_1 \Phi^{--} s + c_2 s \Phi^{++} \label{16}
\end{eqnarray}
with the shorthand notation
\(
s = \vec{\sigma}\vec{p}\,,\;\;
c_i = \omega_i/(\omega_1 m_2 + \omega_2 m_1)
\).
We thus find that \(\Phi\) can be written as
\begin{equation}
\Phi = \hat{\Phi}\,(\Phi^{++},\Phi^{--}) \label{16p}
\end{equation}
with \(\hat{\Phi}\) being a bilinear function.
One can interpret \(\Phi^{++}\) as the upper component and
\(\Phi^{--}\) as the lower component of \(\Phi\), as can be seen
in the nonrelativistic limit where \(\Phi^{--}\) vanishes for solutions
that fulfill \(M \approx m_1+m_2\) and where \(\Phi^{++}\,i\sigma_2\)
becomes the usual Schr\"odinger wave function.

For further discussion it is useful to rewrite the Salpeter
equation in the form of an eigenvalue problem for the bound state
mass \(M\). We follow the treatment of Laga\"e \cite{La} and define
\begin{eqnarray}
\psi(\vec{p}) &:=& \Phi(\vec{p})\,\gamma^0 \label{17}\\
 {} [W(\vec{p},\vec{p}\,')\, \psi(\vec{p}\,')] &:=&
\gamma^0\,[V(\vec{p},\vec{p}\,')\, \Phi(\vec{p}\,')] \label{18}
\end{eqnarray}
The Salpeter equation can now be written as
\begin{equation}
({\cal H}\psi)(\vec{p}) = M\,\psi(\vec{p})
\label{19} \end{equation}
where
\begin{eqnarray}
 ({\cal H}\psi)(\vec{p})
& = & H_1(\vec{p}) \psi(\vec{p}) - \psi(\vec{p}) H_2(\vec{p}) \nonumber \\ &-&
 \int \frac{d^3p'}{(2\pi)^3}  \,
 \Lambda_1^+(\vec{p}) \,[W(\vec{p},\vec{p}\,')\, \psi(\vec{p}\,')]\,
\Lambda_2^-(\vec{p}) \nonumber \\
 &+&\int \frac{d^3p'}{(2\pi)^3}  \,
 \Lambda_1^-(\vec{p}) \,[W(\vec{p},\vec{p}\,')\, \psi(\vec{p}\,')]\,
\Lambda_2^+(\vec{p}) \label{20}
\end{eqnarray}
The equivalence of eq.(\ref{19}) and eq.(\ref{9}) can be shown by applying
the projectors \(\Lambda_i^{\pm}\) to both equations from both sides.
{}From eq.(\ref{19}) one obtains e.g.
\begin{eqnarray}
\Lambda_1^+(\vec{p}) \, \psi(\vec{p}) \, \Lambda_2^+(\vec{p}) &=& 0
\nonumber \\
\Lambda_1^-(\vec{p}) \, \psi(\vec{p}) \, \Lambda_2^-(\vec{p}) &=& 0
 \label{20p}
\end{eqnarray}
as in eq.(\ref{15}) due to the relation
\(\Lambda_i^{\pm}(\vec{p})\,\gamma^0 =
  \gamma^0\,\Lambda_i^{\pm}(-\vec{p})\).
Note that eq.(\ref{20p}) can be also written in the concise form
\begin{equation}
\frac{H_1}{\omega_1}\psi + \psi\frac{H_2}{\omega_2} = 0 \label{21}
\end{equation}

\subsection{Normalization condition and scalar product} \label{IIB}
The normalization for general BS-amplitudes has been given
for bound states with conserved quantum numbers by Nishijima \cite{Ni}
and Mandelstam \cite{Ma}. We follow Cutcosky \cite{Cu}, who treated the
more general case where no current has to be conserved. As this has
already been treated within textbooks (e.g. \cite{Lu}), we will only
give the result for the normalization.
Let the bound state be normalized as
$          \left\langle P\right.\left|P'\right\rangle
               = (2\pi )^3\, 2P^0 \,
                                   \delta^3(\vec{P}-\vec{P'})$.
Then the normalization condition
in graphical representation is given by Fig.\ref{vnorm}
(compare e.g. \cite{LS1}).
Contracting with the momentum of the bound state, this reads explicitly:
\begin{eqnarray}
\label{nrm}
   \int \!\frac{d^4p}{(2\pi)^4} \frac{d^4p'}{(2\pi)^4} \,
                 tr \, \Bigg[  \bar{\chi}_P(p)P^{\mu}\frac{d}{dP^{\mu}}
          \Bigg( I(P,p,p') & &  \\
                +\, i\, K(P,p,p')\Bigg)
                                       \chi_P(p')\Bigg]
                                                    & = & 2iM^2
\nonumber
\end{eqnarray}
The summation over color indices is suppressed here.
$I$ denotes the product of the inverse quark-propagators:
\begin{eqnarray}
 I(P,p,p')_{\alpha\alpha',\beta\beta'}
       & = & \delta^{(4)}(p-p')\,(2\pi)^4   \\ & &
 {S^F_1}_{\alpha\alpha'}^{-1}(\eta_1P+p)\,
 {S^F_2}_{\beta'\beta}^{-1}(-\eta_2P+p) \nonumber
\end{eqnarray}
Note that the vectorial condition of Fig.(\ref{vnorm})
and the scalar normalization (\ref{nrm}) are in fact equivalent which
follows from: (a) the formal covariance of the
equation and (b) the fact that in the rest frame the time component of
Fig.(\ref{vnorm}) gives eq.(\ref{nrm}) and the space components vanish,
as the derivative $d/(dP^i)[I+K]$ is proportional to $p^i,{p'}^i$ or
$\gamma^i$ so that the integrals or the trace on the rhs. of
eq.(\ref{nrm}) give zero.
For an interaction-kernel, which is instantaneous in the rest frame,
i.e. of the type of eq.(\ref{7}), we have:
\begin{eqnarray}
          P^{\mu}\frac{d}{dP^{\mu}} V\left(
                      p_{\perp},p'_{\perp}\right)=0
\end{eqnarray}
so that the contributions of the interaction kernel to the normalization
vanish. At this point we would
like to mention that the BS-equation and the
normalization condition for an instantaneous interaction may
be formulated covariantly, so that the corresponding amplitudes
$\chi$ are correctly normalized in any frame. The explicit
normalization for the corresponding Salpeter amplitudes
$\Phi$ \cite{Ma,Sa}  will be performed in the rest frame.
First we define the vertex functions:
\begin{eqnarray}
     \Gamma_P(p)  &:=&
[S^F_1(p_1)]^{-1} \,\chi_P(p)\;[S^F_2(-p_2)]^{-1} \nonumber \\
\bar{\Gamma}_P(p) &:=&
[S^F_2(-p_2)]^{-1} \,\bar{\chi}_P(p) \;[S^F_1(p_1)]^{-1}
\label{dv}
\end{eqnarray}
With the BS equation (\ref{4}) we then obtain the following important result:
\begin{eqnarray}
  \Gamma_P(p)\left|_{_{P=(M,\vec{0}\,)}}\right.  =
\Gamma(\vec{p}\,) =
  -i\! \int\!\! \frac{d^3p}{(2\pi)^4}
  \left[ V(\vec{p},\vec{p}\,')\Phi(\vec{p}\,')\right]
\label{vert}
\end{eqnarray}
i.e. the vertex-function depends only on the relative three-momentum $\vec{p}$.
This formula allows the reconstruction of the vertex function $\Gamma$
and therefore of the full BS-amplitude $\chi$ from the
Salpeter amplitude $\Phi$. Inserting eq.(\ref{dv}) into the
normalization condition (\ref{nrm}) the dependence on $p^0$ is
completely determined by the quark-propagators, so that the
$p^0$-integration may be performed analytically.

We use the general relation between the BS-amplitude  $\chi$ and it's
adjoint $\bar{\chi}$
for spin-1/2-fermions (see e.g. \cite{IZ} for the scalar case):
\begin{eqnarray}
        \chi(p) & = & -\frac{1}{2\pi i} \int dq^0 \left(
                        \frac{f(q^0,\vec{p}\,)}{p^0-q^0+i\epsilon}
                        +\frac{g(q^0,\vec{p}\,)}{p^0-q^0-i\epsilon}\right)
\\
        \bar{\chi}(p) & = & -\frac{1}{2\pi i} \gamma^0\int dq^0 \left(
\frac{f^{\dagger}(q^0,\vec{p}\,)}{p^0-q^0+i\epsilon}
+\frac{g^{\dagger}(q^0,\vec{p}\,)}{p^0-q^0-i\epsilon}\right)
                                                                \gamma^0
\nonumber
\end{eqnarray}
with matrix valued functions $f$ and $g$.
{}From this we derive the following relations in the special case of an
instantaneous interaction:
\begin{eqnarray}
              \bar{\Gamma}(\vec{p}\,)=
                   - \gamma_0\, \Gamma^{\dagger}(\vec{p}\,)\, \gamma_0
 \hspace{1cm}
              \bar{\Phi}(\vec{p}\,)=
                   \; \; \gamma_0\, \Phi^{\dagger}(\vec{p}\,)\, \gamma_0
\end{eqnarray}
This leads to the normalization condition for the Salpeter-amplitudes in
the rest frame:
\begin{eqnarray}
       \int \frac{d^3p}{(2\pi )^3}
         \, tr \, \Big\{ \Phi^{\dagger}(\vec{p}\,) \Lambda_1^+(\vec{p}\,)
                          \Phi(\vec{p}\,)   \Lambda_2^-(-\vec{p}\,) & &
\label{n1}\\    -         \Phi^{\dagger}(\vec{p}\,) \Lambda_1^-(\vec{p}\,)
                          \Phi(\vec{p}\,)   \Lambda_2^+(-\vec{p}\,)
                   & \Big\}& =   (2\pi )^2 \, 2M
\nonumber
\end{eqnarray}
It also may be expressed in terms of the 2$\times$2 amplitudes
$\Phi^{++}$ and $\Phi^{--}$ defined in eq.(\ref{14}) as:
\begin{eqnarray}
 \int \!\frac{d^3p}{(2\pi )^3}
        \frac{2 \omega_1 \omega_2}{\omega_1 m_2 + \omega_2 m_1}
  \, tr \, \Bigg\{\;(\Phi^{++}(\vec{p}\,))^{\dagger}  \,\Phi^{++}(\vec{p}\,)
&& \label{norm} \\  -\, (\Phi^{--}(\vec{p}\,))^{\dagger} \,
\Phi^{--}(\vec{p}\,)\;
          \Bigg\}            = (2\pi)^2& 2&M
\nonumber
\end{eqnarray}
This form also shows the connection to the nonrelativistic
norm: in the NR limit
$\Phi^{++}(\vec{p}\,)\,i\sigma_2$ becomes the usual Schr\"odinger wave
function and $\Phi^{--}(\vec{p}\,)$ goes to zero as ${\vec{p}\,}^2/m^2\cdot
\Phi^{++}(\vec{p}\,)$. Furthermore the weight function becomes
equal to unity so that we obtain the usual Schr\"odinger
normalization. For deeply bound states however we have
appreciable deviations from this norm, as the lower amplitude $\Phi^{--}$
is of the same order as the upper component $\Phi^{++}$ (see Sec.\ref{IIC}).

Eq.(\ref{n1}) motivates the definition of a
scalar product for amplitudes \(\psi_1=\Phi_1\gamma^0\)
and \(\psi_2=\Phi_2\gamma^0\) as
\begin{eqnarray}
\left\langle \psi_1 \right.\left| \psi_2 \right\rangle
&=& \int\,\mbox{tr}\,\left(
    \psi_1^{\dagger}\,\Lambda_1^+\,
    \psi_2\,\Lambda_2^- -
    \psi_1^{\dagger}\,\Lambda_1^-\,
    \psi_2\,\Lambda_2^+ \right) \nonumber \\
&=& \frac{1}{2}\,\int\,\mbox{tr}\,
\left[ \psi_1^{\dagger}\,\left(\frac{H_1}{\omega_1} \,\psi_2
 - \psi_2\,\frac{H_2}{\omega_2}\right) \right]
\label{22}
\end{eqnarray}
with all quantities depending on \(\vec{p}\) and the notation
\(\int=\int\,d^3p/(2\pi)^3\). Note that this scalar product is not
positive definite. The normalization condition (\ref{n1}) for solutions of the
Salpeter equation is then given as
\begin{equation}
\left\langle \psi \right.\left| \psi \right\rangle = (2\pi)^2\,2M
\end{equation}
The following discussion will be restricted to amplitudes satisfying
eq.(\ref{21}). In that case one has
\begin{eqnarray}
\left\langle \psi_1 \right.\left|
\,{\cal H}\, \psi_2 \right\rangle
&=& \int\,(\omega_1+\omega_2)\,\mbox{tr}\,
\left(\psi_1^{\dagger}\,\psi_2\right) \nonumber \\
&-& \int\,\int'\, \mbox{tr}\,
\left(\psi^{\dagger}_1\,W\psi_2'\right) \label{21p}
\end{eqnarray}
where the prime indicates the dependence of \(\psi_2\) on \(\vec{p}\,'\).
If one considers kernels that fulfill
\(
\int\int' \mbox{tr}\,(\psi^{\dagger}_1\,W\psi_2') =
\int\int' \mbox{tr}\,(\psi^{\dagger}_2\,W\psi_1')^*
\),
which is valid for a wide class of interactions
(e.g. for \(W\psi'=f\left((\vec{p}-\vec{p}\,')^2\right)\,
\Gamma_1\,\psi(\vec{p}\,')\,\Gamma_2\) with hermitian matrices
\(\Gamma_i\) and a scalar function \(f\)),
the Salpeter hamiltonian \({\cal H}\) is selfadjoint
with respect to the scalar product given in eq.(\ref{22}), i.e.
\begin{equation}
\left\langle \psi_1 \right.\left|\,{\cal H}\, \psi_2 \right\rangle =
\left\langle \,{\cal H}\,\psi_1 \right.\left|\,\psi_2 \right\rangle
\end{equation}
which has two important consequences, namely
\begin{itemize}
\item
bound state masses \(M\) are real for eigen functions \(\psi\)
with nonzero norm
\(\left\langle \psi \right.\left|\,\psi \right\rangle \not= 0\)
\item
amplitudes \(\psi_1\) and \(\psi_2\) corresponding to different
eigenvalues \(M_1 \not= M_2^*\) are orthogonal, i.e.
\(\left\langle \psi_1 \right.\left|\,\psi_2 \right\rangle = 0\)
\end{itemize}
The first point can be seen easily from
\(
M\,\left\langle \psi \right.\left|\,\psi \right\rangle =
\left\langle \psi \right.\left|{\cal H}\,\,\psi \right\rangle =
\left\langle {\cal H}\,\psi \right.\left|\,\psi \right\rangle =
M^*\,\left\langle \psi \right.\left|\,\psi \right\rangle
\)
and the second point follows from
\(
M_2\,\left\langle \psi_1 \right.\left|\,\psi_2 \right\rangle =
\left\langle \psi_1 \right.\left|{\cal H}\,\,\psi_2 \right\rangle =
\left\langle {\cal H}\,\psi_1 \right.\left|\,\psi_2 \right\rangle =
M_1^*\,\left\langle \psi_1 \right.\left|\,\psi_2 \right\rangle
\).

\subsection{Structure of the Solutions} \label{IIC}
The Salpeter equation exhibits some further general structures
connecting solutions with positive and negative eigenvalues.
For the case of equal quark masses
J.F. Laga\"e \cite{La} has shown that for kernels
satisfying \((W\psi)^{\dagger}=W\psi^{\dagger}\) the eigenvalues will
come in
pairs of opposite sign, the corresponding eigen functions having
normalizations with opposite sign. In the following we will extend
this result to the general case of unequal quark masses and
compare the block structure of the conjugate solutions.
We further show that nondegenerate bound states with mass
\(M=0\) have zero norm
\(\left\langle \psi \right.\left|\,\psi \right\rangle = 0\).
The discussion of physical acceptable solutions is postponed
to the end of this section.

To deal with the unequal mass case we first investigate
the structure of the BS-equation under charge conjugation.
The details are shown in the appendix with the result
that solutions of
\begin{eqnarray}
({\cal H}_{f_1f_2}\psi_{f_1f_2})(\vec{p}) = M\,\psi_{f_1f_2}(\vec{p})
\nonumber\\
({\cal H}_{f_2f_1}\psi_{f_2f_1})(\vec{p}) = M\,\psi_{f_2f_1}(\vec{p})
\label{31}
\end{eqnarray}
are related through
\begin{equation}
\psi_{f_1f_2}(\vec{p}) = -S_C\;\,{}^t \psi_{f_2f_1}(-\vec{p})\,S_C
\label{32} \end{equation}
with \(S_C\) given in eq.(\ref{sc}) in the appendix. The indices of
\({\cal H}\)
denote the flavor dependence of \(H_i\) and \(\Lambda_i^{\pm}\)
in eq.(\ref{20}). For simplicity only the case without flavor mixing is
considered, the generalization being straightforward.

In the following discussion we will assume that the BS-kernel
satisfies
\( (W_{f_1f_2}\psi'_{f_1f_2})^{\dagger} =
W_{f_2f_1}(\psi'_{f_1f_2})^{\dagger} \) which is fulfilled e.g. for kernels
of the form
\(W_{f_1f_2}\psi'_{f_1f_2} = f\left((\vec{p}-\vec{p}\,')^2\right)\,
\Gamma\,\psi_{f_1f_2}(\vec{p}\,')\,\Gamma\) with a hermitian matrix
\(\Gamma\). The hermitian conjugate of eq.(\ref{19}) thus leads to
\begin{equation}
  -({\cal H}_{f_1f_2}\psi_{f_1f_2})^{\dagger}
 =  {\cal H}_{f_2f_1}\psi_{f_1f_2}^{\dagger}
 = -M^*\,\psi_{f_1f_2}^{\dagger} \label{33}
\end{equation}
Renaming \(f_1 \leftrightarrow f_2\) and comparing this equation to
eq.(\ref{31}) we thus have shown that
\begin{itemize}
\item
solutions of the Salpeter equation come in pairs
\((\psi_{f_1f_2},\,M)\) and \((\psi_{f_2f_1}^{\dagger},\,-M^*)\)
where eq.(\ref{32}) connects the two solutions.
\end{itemize}
Consider the normalization of \(\psi_{f_1f_2}\) and
\(\psi_{f_2f_1}^{\dagger}\):
With eq.(\ref{32}) and the relation
\(S_C\,\Lambda_i^{\pm}(\vec{p})\,S_C = {}^t\Lambda_i^{\mp}(-\vec{p})\)
one finds
\(
\left\langle \psi_{f_2f_1} \right.\left|\,\psi_{f_2f_1}
\right\rangle_{f_2f_1} =
\left\langle \psi_{f_1f_2} \right.\left|\,\psi_{f_1f_2}
\right\rangle_{f_1f_2} \)
where the indices of the scalar product determine the flavors of
\(\Lambda_1^{\pm}\) and \(\Lambda_2^{\pm}\) in eq.(\ref{22}).
On the other hand cyclic permutation under the trace shows that
\(
 \langle \psi_{f_2f_1}^A |
\,\psi_{f_2f_1}^B \rangle_{f_2f_1} =
-\langle (\psi_{f_2f_1}^B)^{\dagger} |
\,(\psi_{f_2f_1}^A)^{\dagger} \rangle_{f_1f_2}
\)
so that we have
\begin{equation}
 \langle \psi_{f_1f_2} |
\,\psi_{f_1f_2} \rangle_{f_1f_2} =
-\langle \psi_{f_2f_1}^{\dagger} |
\,\psi_{f_2f_1}^{\dagger} \rangle_{f_1f_2}
\label{36} \end{equation}
and we find that
\begin{itemize}
\item
the states with eigenvalues \(M\) and \(-M^*\) have opposite norm.
\end{itemize}

Let us now compare the block matrix structure of the two conjugated
solutions. With the angular momentum decomposition
eq.(\ref{120}) (see appendix)
\begin{eqnarray}
\Phi^{++}(\vec{p}) &=&
 \sum_{L\,S}\,{\cal R}_{LS}^{(+)}(p) \,
 \left[Y_L(\Omega_p) \otimes \varphi_S \right]^J \nonumber \\
\Phi^{--}(\vec{p}) &=&
 \sum_{L\,S}\,{\cal R}_{LS}^{(-)}(p) \,
 \left[Y_L(\Omega_p) \otimes \varphi_S \right]^J
\end{eqnarray}
with the 2\(\times\)2-matrices
\(\varphi_{0\,0} = 1/\sqrt{2},\;
\varphi_{1\,q}   = \sigma_q/\sqrt{2}\)
it is straightforward to show that
\begin{eqnarray}
\lefteqn{
\left[\left(\Phi_{f_2f_1}^{++}\right)^{JM_J}\right]^{\dagger}
= (-1)^{-J-M_J} } \nonumber \\
&&\sum_{LS}\,(-1)^{L+S}\,
\left[{\cal R}_{LS}^{(+)}(p)\right]_{f_2f_1}\,
\left[Y_L(\Omega_p) \otimes \varphi_S \right]^J_{-M_J}
\end{eqnarray}
and similarly for \(\Phi^{--}\). It
is shown in the appendix that
\(
[{\cal R}_{LS}^{(+)}(p)]_{f_2f_1} = (-1)^{L+S}\,
[{\cal R}_{LS}^{(+)}(p)]_{f_1f_2}
\).
Since \(L\) and \(S\) are integer the phase vanishes and we obtain
the result
\begin{equation}
\left[\left(\Phi_{f_2f_1}^{++}\right)^{JM_J}\right]^{\dagger}
= (-1)^{-J-M_J}
\left(\Phi_{f_1f_2}^{++}\right)^{J\,-M_J} \label{49}
\end{equation}
According to eqs.(\ref{16p}),(\ref{17}) we write
\begin{equation}
\psi_{f_2f_1}^{\dagger} = \gamma^0\,
\left[ \hat{\Phi}_{f_2f_1}(\,\phi^{++}_{f_2f_1},\,\phi^{--}_{f_2f_1})
\right]^{\dagger} \label{51}
\end{equation}
where the indices of \(\hat{\Phi}\) indicate the flavor dependence of
\(c_i\) in eq.(\ref{16}).
The hermitian conjugate of eq.(\ref{20p}) with \(f_1\) and \(f_2\)
interchanged gives
\(
\Lambda_{f_1}^+\,\psi_{f_2f_1}^{\dagger}\,\Lambda_{f_2}^+ = 0 \) and \(
\Lambda_{f_1}^-\,\psi_{f_2f_1}^{\dagger}\,\Lambda_{f_2}^- = 0
\)
so that we can write
\begin{equation}
\psi_{f_2f_1}^{\dagger} = \hat{\Phi}_{f_1f_2}
\left(\xi^{++}_{f_1f_2},\,\xi^{--}_{f_1f_2}\right)\,\gamma^0 \label{52}
\end{equation}
with some amplitudes \(\xi^{++},\,\xi^{--}\) that can be determined by
comparing eq.(\ref{52}) to eq.(\ref{51})
with the result
\(
(\xi^{++}_{f_1f_2})^{JM_J}
 =  -[(\Phi^{--}_{f_2f_1})^{JM_J}]^{\dagger}
 =  (-1)^{1-J-M_J}\,(\Phi^{--}_{f_1f_2})^{J\,-M_J}
\)
and the same expression with \(++\) and \(--\) interchanged. According to
eq.(\ref{52}) we thus can write
\begin{eqnarray}
\left(\psi_{f_2f_1}^{JM_J}\right)^{\dagger}& =&
(-1)^{1-J-M_J} \label{60}\\&& \hat{\Phi}_{f_1f_2}
\left[
\left(\Phi^{--}_{f_1f_2}\right)^{J\,-M_J},
\left(\Phi^{++}_{f_1f_2}\right)^{J\,-M_J}
\right]\,\gamma^0  \nonumber
\end{eqnarray}
We thus have the result that
\begin{itemize}
\item
exchanging the functions \(\Phi^{++}\) and \(\Phi^{--}\)
in \(\Phi=\hat{\Phi}\,\left(\Phi^{++},\Phi^{--}\right)\) turns an
amplitude with eigenvalue \(M\) into an amplitude with eigenvalue \(-M^*\).
\end{itemize}

With the relations obtained above it is easy to investigate
the eigenvalue \(M=0\) which is assumed to be
not degenerate apart from the trivial degeneracy in the angular
momentum projection \(M_J\). From eqs.(\ref{31}),(\ref{33}) we have
\begin{equation}
{\cal H}_{f_1f_2}\psi_{f_1f_2}   = 0 \hspace{1cm}
{\cal H}_{f_1f_2}\psi_{f_2f_1}^{\dagger} = 0
\end{equation}
which through eq.(\ref{60}) implies
\(
(\psi_{f_2f_1}^{JM_J})^{\dagger} = \lambda\,\psi_{f_1f_2}^{J\,-M_J}
\)
with \( |\lambda| = 1\). Then eq.(\ref{36}) gives
\begin{eqnarray}
 \left\langle \psi_{f_1f_2}^{JM_J} \right.\left|
\,\psi_{f_1f_2}^{JM_J} \right\rangle_{f_1f_2} &=&
-\left\langle \left(\psi_{f_2f_1}^{JM_J}\right)^{\dagger} \right.\left|
\,\left(\psi_{f_2f_1}^{JM_J}\right)^{\dagger} \right\rangle_{f_1f_2}
\nonumber \\
&=& - \left\langle \psi_{f_1f_2}^{J\,-M_J} \right.\left|
\,\psi_{f_1f_2}^{J\,-M_J} \right\rangle_{f_1f_2}
\end{eqnarray}
Since the scalar product is invariant under rotations
one can substitute \(-M_J\) by \(M_J\) and obtains
\begin{equation}
 \left\langle \psi_{f_1f_2}^{JM_J} \right.\left|
\,\psi_{f_1f_2}^{JM_J} \right\rangle_{f_1f_2} = 0
\end{equation}
So we find that
\begin{itemize}
\item
nondegenerate eigen functions with eigenvalue \(M=0\) have zero norm.
\end{itemize}
{}From eq.(\ref{60}) it is clear that
\((\psi_{f_2f_1}^{JM_J})^{\dagger} = \lambda\,\psi_{f_1f_2}^{J\,-M_J}\)
is equivalent to setting
\begin{equation}
\Phi^{++} = \pm \Phi^{--}
\end{equation}
This equation illustrates a common aspect of the Salpeter equation:
in the nonrelativistic limit with \(M \approx m_1+m_2\)
the large component \(\Phi^{++}\) dominates
over the small component \(\Phi^{--}\), but if one goes to deeply
bound states (e.g. by increasing the coupling constant of an attractive
interaction) the two components become more and more equal until
finally \(\Phi^{++} = \pm \Phi^{--}\) is achieved for \(M=0\).

{}From the discussion above it has become clear that we have to
identify the physically acceptable solutions.
There are two criteria making sure that
a solution is acceptable:
\begin{itemize}
\item
The norm of the solution has to be nonzero
which automatically implies that \(M\) is real.
\item
The eigenvalue \(M\) and the norm have to be positive
in order to fulfill the normalization condition
\(  \left\langle \psi \right.\left|\,\psi \right\rangle
 = (2\pi)^2\,2M\).
\end{itemize}
We would like to mention that the typical doubling
of the physical eigenvalues is well known from the
RPA equations in nonrelativistic many particle theory \cite{RS}.
For a hamiltonian \(H\) with spectrum \(E_n\)
the RPA equations have solutions \(E_n,\,-E_n\).
This doubling can be traced back to the appearance of the
time ordering operator \(T\) in the definition of
the particle-hole propagator. Therefore neglecting
negative mass eigenvalues is consistent with the RPA
structure of the Salpeter equation (compare chap.\ref{III}.

The role of the solutions with \(M=0\) and
\(  \left\langle \psi \right.\left|\,\psi \right\rangle = 0\)
is not clear.
On one hand there is a priori no contradiction with the normalization
condition. On the other hand the Salpeter equation has been obtained
in the rest frame of the bound state, i.e. one first performs the
limit \(\vec{P} \rightarrow 0\) and then investigates the case
\(M \rightarrow 0\). However, the correct procedure for massless bound states
is first to perform the limit \(M \rightarrow 0\) in the BS-equation.
In the resulting equation one then can study
the limit \(\vec{P} \rightarrow 0\). It cannot generally be expected
that exchanging the limits for \(\vec{P}\) and \(M\) leads to
equivalent results (compare \cite{LS2} for a more detailed discussion
of this problem). Furthermore the definition of the instantaneous
interaction kernel eq.(\ref{7}) becomes dubious since \(p_{\perp}\)
is not well defined for \(P \rightarrow 0\). We therefore prefer
to require \(M>0\) and
\(  \left\langle \psi \right.\left|\,\psi \right\rangle > 0\) for
physically acceptable solutions.

\section{Numerical treatment} \label{III}
{}From the definition of \(\hat{\Phi}\) in eq.(\ref{16p}) it is easy
to derive a basis expansion for \(\psi=\Phi\gamma^0\). Let
\begin{equation}
 E_i(\vec{p}) = R_{n_i L_i}(p)\,
 \left[ Y_{L_i}(\Omega_p) \otimes \varphi_{S_i} \right]^J_{M_J} \label{80}
\end{equation}
be a complete set of 2\(\times\)2 basis functions with real
radial functions \(R_{n_i L_i}(p)\).
The basis functions are chosen orthonormal with respect to
the usual scalar product given by
\begin{equation}
 \left( E_i \right.\left|\,E_j \right) =
\int \frac{d^3p}{(2\pi)^3}\,\mbox{tr}\,
\left[E_i^+(\vec{p})\,E_j(\vec{p})\right]
= \delta_{ij} \label{81}
\end{equation}
where the trace just gives the usual scalar product for the spin
matrices \(\mbox{tr}\,\varphi^+_{SM_S}\varphi_{S'M_S'} =
\delta_{SS'}\delta_{M_SM_S'} \). Note that the angular structure
of the basis functions matches the structure of \(\Phi^{++}\) and
\(\Phi^{--}\) as given in eq.(\ref{120}). We choose \(R_{n_iL_i}(p)\)
to be real functions.
It is now possible to expand
\begin{eqnarray}
\Phi^{++}(\vec{p}) &=& \sum_{i=1}^{\infty}\,a_i^{(+)}\,E_i(\vec{p}) \\
\Phi^{--}(\vec{p}) &=& \sum_{i=1}^{\infty}\,a_i^{(-)}\,E_i(\vec{p})
\label{82}
\end{eqnarray}
which implies \({\cal R}^{(\pm)}_{LS}(p) =
\sum_{i=1}^{\infty}\,a_i^{(\pm)}\,R_{n_iL_i}(p)\,\delta_{LL_i}\,\delta_{SS_i}
\)
for the radial wave function. Since \({\cal R}^{(\pm)}_{LS}(p)\)
are real functions in most cases of interest, the coefficients
\(a_i^{(\pm)}\) then also have to be real.
Now define the 4\(\times\)4-amplitudes
\begin{eqnarray}
e_i^{(+)} &=& \hat{\Phi}\,(E_i,0)\,\gamma^0
\nonumber \\
e_i^{(-)} &=& \hat{\Phi}\,(0,E_i)\,\gamma^0  \label{83}
\end{eqnarray}
Note that these functions are not orthogonal
with respect to the scalar product given in eq.(\ref{22}).
Since \(\hat{\Phi}\) is bilinear we nevertheless can
expand \(\psi\) as
\begin{equation}
\psi = \sum_{i=1}^{\infty}\,
\left(a_i^{(+)}\,e_i^{(+)} + a_i^{(-)}\,e_i^{(-)} \right)
\label{84}
\end{equation}
so that the constraint
\(\Lambda_1^+ \psi \Lambda_2^+ = \Lambda_1^- \psi \Lambda_2^- = 0\)
is automatically fulfilled.
The Salpeter equation \( {\cal H}\psi = M\psi \)
can now be written as the matrix equation
\begin{eqnarray}
\left( \!\! \begin{array}{*{2}{c}}
H^{++} & H^{+-} \\
H^{-+} & H^{--}
\end{array} \!\! \right)\!
\left( \!\! \begin{array}{*{1}{c}}
a^{(+)} \\ a^{(-)}
\end{array} \!\! \right) = M
\left( \!\! \begin{array}{*{2}{c}}
N^{++} & N^{+-} \\
N^{-+} & N^{--}
\end{array} \!\! \right)\!
\left( \!\! \begin{array}{*{1}{c}}
a^{(+)} \\ a^{(-)}
\end{array} \!\! \right) \label{85}
\end{eqnarray}
with
\(H^{ss'}_{ij} = \langle e_i^{(s)} | {\cal H}\,e_j^{(s')} \rangle \)
and
\(N^{ss'}_{ij} = \langle e_i^{(s)} | e_j^{(s')} \rangle \).
{}From the definition of the scalar product one easily sees that
\(N^{++}=-N^{--}\) and \(N^{+-}=N^{-+}=0\). Furthermore we find
from eqs.(\ref{60}),(\ref{83}) that \(e_i^{(+)}\) and \(e_i^{(-)}\)
are connected by
\( [(e_i^{(+)})^{JM_J}_{f_1f_2}]^{\dagger} = (-1)^{1-J-M_J}
    (e_i^{(-)})^{J\,-M_J}_{f_2f_2} \)
so that we can use eqs.(\ref{33}),(\ref{36}) as well as
the invariance of the scalar product under flavor exchange
and under the replacement \(-M_J \rightarrow M_J\) to obtain
\( H^{--}_{ij}=(H^{++}_{ij})^*\) and \(H^{-+}_{ij}=(H^{+-}_{ij})^*\).
The matrix representation of the Salpeter equation thus takes the form
\begin{eqnarray}
\lefteqn{ \left( \!\!\begin{array}{*{2}{c}}
H^{++} & H^{+-} \\
(H^{+-})^* & (H^{++})^*
\end{array} \!\!\right)
\left( \!\!\begin{array}{*{1}{c}}
a^{(+)} \\ a^{(-)}
\end{array} \!\!\right) = } \nonumber \\ &=& M
\left( \!\!\begin{array}{*{2}{c}}
N^{++} & 0 \\
0      & -N^{++}
\end{array} \!\!\right)
\left( \!\!\begin{array}{*{1}{c}}
a^{(+)} \\ a^{(-)}
\end{array} \!\!\right) \label{87}
\end{eqnarray}
which is of the same form as the well-known RPA equations
in nuclear physics \cite{RS}. Let \((a^{(+)},a^{(-)})\)
be an eigenvector with eigenvalue \(M\). Then eq.(\ref{87})
shows that \(((a^{(-)})^*,(a^{(+)})^*)\) is an eigenvector with
eigenvalue \(-M^*\) which is just the result of the previous
section.
Usually the Salpeter hamiltonian \({\cal H}\) has the property
that the matrix elements \(H_{ij}^{ss'}\)
and also the eigenvector coefficients \(a_i^{(\pm)}\) are real within
the basis given above. In that case and since
\(N_{ij}^{ss'}\) is real, also \(M\) must be real.
This result has already been shown before for
eigen vectors with nonzero norm.

Furthermore we see that if
\(M=0\) is an eigenvalue we expect from eq.(\ref{87})
that the eigenvector fulfills \(a^{(+)}=\pm a^{(-)}\)
and we reobtain the result that this solution has zero norm.

In a numerical treatment only a finite basis \(i \le i_{max} \approx 10\)
can be taken into account. Then eq.(\ref{87}) becomes a finite
matrix equation that can be solved with standard numerical methods.
One thus obtains an approximate eigenvalue \(M_{\beta}\) and an approximate
eigen function \(\psi_{\beta}\) to the Salpeter equation
that exactly fulfill the relation
\begin{equation}
\left\langle \psi_{\beta} \right. \left|
{\cal H}\,\psi_{\beta} \right\rangle = M_{\beta}\,
\left\langle \psi_{\beta} \right. \left|
\psi_{\beta} \right\rangle  \label{88}
\end{equation}
The index \(\beta\) indicates that the basis states \(E_i\) and
thus \(\psi_{\beta}\) depend on a variational parameter \(\beta\)
of dimension \(MeV^{-1}\) that sets the absolute scale for the
momentum dependence via
\(E_i^{\beta}(\vec{p}) = \beta^{3/2}\,E_i^{\beta=1}(\vec{p}\beta) \).

If solutions of nonzero norm are considered
the Salpeter equation is equivalent to the variational problem \cite{La}
\begin{equation}
\delta M[\psi] = \delta \frac{
\left\langle \psi \right. \left|
{\cal H}\,\psi \right\rangle} {
\left\langle \psi \right. \left|
\psi \right\rangle} = 0  \label{90}
\end{equation}
where the variation \(\delta\) is taken over all functions \(\psi\)
with nonzero norm that fulfill
\(\Lambda_1^+ \psi \Lambda_2^+ = \Lambda_1^- \psi \Lambda_2^- = 0\).
According to eq.(\ref{88}) we make the
variational ansatz \(\psi=\psi_{\beta}\) implying
\(M[\psi_{\beta}]=M_{\beta}\) and look for stationary points
of \(M_{\beta}\) as a function of \(\beta\) (which also
fixes \(\beta\) for each meson).

The calculation of the matrix elements within an explicit
model will be postponed to our second paper. At this point
we would only like to make a few technical comments.
The matrix elements of the
interaction kernel can be efficiently calculated by inserting
two complete sets of basis functions written schematically as
\begin{eqnarray}
\lefteqn{
\left\langle i \left|
\,f_1(\vec{p})\,V(r)\,f_2(\vec{p}\,')\,
\right| j \right\rangle} \nonumber \\ &=& \sum_{g,h}\,
\left\langle i \left|
f_1(\vec{p})
\right| g \right\rangle \,
\left\langle g \left|
V(r)
\right| h \right\rangle \,
\left\langle h \left|
f_2(\vec{p}\,')
\right| j \right\rangle
\end{eqnarray}
so that \(V(r)\) can be parameterized in coordinate space.
A suitable choice for the basis functions is given by the functions
\(R_{nL}(y)=N_{nL}\,y^L\,L_n^{2L+2}(y)\,e^{-y/2}\) with
\(y=p\beta\) and \(L_n^{2L+2}(y)\) being a Laguerre polynomial.
We found that about ten basis states are sufficient to solve
the Salpeter equation with rather high accuracy.
The choice of \(3\)-dimensional harmonic oscillator functions
is less favored since their asymptotic behavior
\(\sim e^{-y^2/2}\) for \(y\rightarrow\infty\) turns out
to be not appropriate for our quark model (especially for deeply bound
states like the Pion).

\section{Decay observables} \label{IV}
Apart from describing the mass spectrum of mesons, any realistic
model must also be able to describe mesonic transitions and decays.
The important question arises whether a good description of
the extremely deep bound states as pion or kaon can be combined with a
reasonable description of confinement.
The BS formalism offers a natural framework, as the role of the
lower component of the wave function turns
out to be crucial for the correct normalization and calculation of the
decays. This can be seen most clearly in the following formulas for
the leptonic decays and the weak decay constants.

We furthermore show how to reconstruct the full BS amplitude, which
may be used as a starting point for the calculation of electromagnetic
or hadronic transitions.

\subsection{Leptonic decay width and weak decay constant}
The transition or decay of bound states are calculated from BS
amplitudes using the formalism given by
Mandelstam \cite{Ma}. We will merely sketch it by considering first
the leptonic decays of vector mesons.
The corresponding Feynman diagram is given by Fig.\ref{mandel}
where $K_{lep}$ denotes the kernel which is irreducible with respect to the
incoming $q\bar{q}$-pair and the outgoing $l^+l^-$-pair.
If we consider only graphs of leading order in the electromagnetic coupling
constant we obtain the approximation on the right hand side.

The exceptional role of these decays (together with the weak decay
constants) is that if the BS kernel would be exact also the
decay amplitudes would be correct to any
order in the strong interaction. The hadronic part of the
transition matrix element is given by \cite{Lu}:
\begin{eqnarray}
      \left\langle\,0\,\left|\,j_{\mu}(0)\,
            \right|\,P\, 1^-\right\rangle \left|_{_{P=(M,\vec{0})}}\right.
              &=&  - tr \left( \gamma_{\mu}\, \chi_{_{(M,\vec{0})}}(x=0)\right)
\nonumber \\
              &=&  - \int \!\! \frac{d^3p}{(2\pi)^{4}}
                  \, tr \,\left( \gamma_{\mu}\, \Phi(\vec{p})\right)
\label{lecur}
\end{eqnarray}
The instantaneous approximation thus simplyfies this
calculation, as one can express the transition in terms of the
Salpeter amplitude $\Phi$. It may be decomposed using the transformation
properties under rotation and parity, as is shown in appendix A
(compare eq.(\ref{adchi})).
For $J^P\,=\,1^-$ mesons it reads (with $p:=|\vec{p}\,|)$:
\begin{eqnarray}
\lefteqn{
       \Phi^{1M}(\vec{{p}}\,)  =  \;\;\;\Phi^{00}_{10}(p)\,Y_{1M}(\Omega_p)
              + \Phi^{01}_{10}(p)\,\gamma_0 \,Y_{1M}(\Omega_p)}
 \\ &&
   + \Phi^{11}_{11}(p)\,\gamma_5\gamma_0
                  \left[ Y_{1}(\Omega_p)\times\gamma\right]_{1M}
      + \Phi^{10}_{11}(p)\,\gamma_5
         \left[ Y_{1}(\Omega_p)\times\gamma\right]_{1M}
\nonumber
\\
  & & + \Phi^{00}_{01}(p)\,\left[ Y_{0}(\Omega_p)\times\gamma\right]_{1M}
      + \Phi^{01}_{01}(p)\,\gamma_0\,
              \left[ Y_{0}(\Omega_p)\times\gamma\right]_{1M}
\nonumber \\ &&
      + \Phi^{00}_{21}(p)\,\left[ Y_{2}(\Omega_p)\times\gamma\right]_{1M}
      + \Phi^{01}_{21}(p)\,\gamma_0\,
                   \left[ Y_{2}(\Omega_p)\times\gamma\right]_{1M}
\nonumber
\end{eqnarray}
The radial amplitudes $\Phi^{a_ig_i}_{l_is_i}$ are determined completely by the
functions $\Phi^{++}$ and $\Phi^{--}$. For a $1^-$ meson the
latter have the form:
\begin{eqnarray}
     \Phi^{\pm\pm}_{1M}(\vec{p}\,) &=&
      {\cal R}_{01}^{(\pm)}(p)
    \,\left[Y_{0}(\Omega_p)\,\times\,\varphi_{1}\right]_{1M} \nonumber \\
    &+& {\cal R}_{21}^{(\pm)}(p)\, \left[
         Y_{2}(\Omega_p)\,\times\,\varphi_{1}\right]_{1M}
\end{eqnarray}
The integration and trace in eq.(\ref{lecur}) for the
current pick up only the s-wave amplitude $\Phi^{00}_{01}$.
The usual spin summation and averaging leads to the decay width:
\begin{eqnarray}
\lefteqn{     \Gamma (1^-\rightarrow l^+l^-) =}\\ & = &
        24 \frac{\alpha^2 \,\tilde{e}_q^2}{M^3}
        \left| \int\! \frac{p^2dp}{(2\pi)^{3} }
                       \left({\cal R}^{(+)}_{01}(p)-{\cal
                         R}^{(-)}_{01}(p)\right)\right|^2
\nonumber
\end{eqnarray}
where $\alpha=1/137$ is the electromagnetic coupling constant and
$\tilde{e}_q$ is the fraction of the quark charge compared to the electron.
Apart from the contribution of the lower component $\Phi^{--}$ to the
normalization eq.(\ref{norm}), its importance comes out here very clearly.
For $M$ we use the experimental meson mass to obtain the correct phase space.

The next observables to be considered are the weak decay constants
$f_{\pi}$ and $f_{_K}$. They are defined by the matrix
element of the axial current \cite{Na} (with this definition
$f_{\pi}^{(exp)}$ = 132 MeV):
\begin{eqnarray}
       i\,f_{\pi}P_{\mu} & = &
      \left\langle\,0\,\left|\,j_{\mu}^5(0)\,
            \right|\,P\,0^-\,\right\rangle
\end{eqnarray}
Again for an instantaneous interaction this can be evaluated from the
Salpeter amplitude
\begin{eqnarray}
     f_{\pi} & = & \left|\;\frac{\sqrt{3}}{M}\int \! \frac{d^{3}p}{(2\pi)^{4}}
            \, tr \,\left(\Phi (\vec{p}\,)\gamma_0\gamma_5\right)\right|
\end{eqnarray}
and in terms of the
$\Phi^{\pm\pm}={\cal R}_{00}^{(\pm)}\,Y_{00}\,\varphi_{00}$:
\begin{eqnarray}
     f_{\pi}& = &\left|\;\frac{\sqrt{3}}{M}\int \! \frac{d^{3}p}{(2\pi)^{4}}
            \, tr \,\left(\Phi^{++} (\vec{p}\,)-\Phi^{--}
            (\vec{p}\,)\right)\right|
\nonumber \\ &=&
            \left|\;\frac{\sqrt{24\pi}}{M}\int \! \frac{p^2dp}{(2\pi)^{4}}
            \left(\,{\cal R}_{00}^{(+)}(p)-{\cal R}_{00}^{(-)}\,\right)\right|
\end{eqnarray}

\subsection{The decay $\pi^0,\eta \rightarrow 2\gamma$}
These decays provide another test for the description of the low lying
pseudoscalar mesons. To our knowledge they have not been calculated in
the framework of the full Salpeter equation
(for a slightly more restricted ansatz see Mitra et al. \cite{Mi}).
The basic idea here is to reconstruct the vertex function $\Gamma$ from the
Salpeter amplitude $\Phi$ by means of the BS-eq.(\ref{vert}) itself, which
gives the full fourdimensional structure of the BS amplitude $\chi$. This
has to be taken into account for a correct description of decays.
The corresponding Feynman diagrams for the neutral pseudoscalar
decay in lowest order of the interaction are given in Fig.\ref{twogamma}.
Of course this is not correct to any order
in the strong interaction like e.g. in the case of the pion decay
constant, as we obviously neglect the strong interaction of the
intermediate quark. Therefore we expect these calculations to be less
accurate.

The T-matrix element in lowest order for a meson with mass $M$ decaying
into two photons with momenta $k_1, k_2$ is given by
\begin{eqnarray}
T & = & -i \sqrt{3} (ie_q)^2 \nonumber
\\ & & \int \!\! \frac{d^4p}{(2\pi)^4}
              \; tr \Bigg\{ \;
        S^F(P/2+p)\; \Gamma  (\vec{p}\,)\; S^F(-P/2+p)
\nonumber \\ &&
       \Bigg[ \varepsilon\!\!\! /_1\; S^F(-P/2+p-k_1)\; \varepsilon\!\!\! /_2
\nonumber \\ &&
        + \varepsilon\!\!\! /_2\; S^F(-P/2+p-k_2)\; \varepsilon\!\!\! /_1
                    \Bigg] \; \Bigg\}
\label{tmat}
\end{eqnarray}
Therefore we have to reconstruct the vertex function $\Gamma$ from the
Salpeter amplitude $\Phi$ by means of eq.(\ref{vert}). Both of them have
an angular decomposition according to eq.(\ref{adchi}). The four scalar
amplitudes depending only on $|\vec{p}\,|$ are expanded in a basis of
radial wave functions. The quark propagator can be written as
\begin{eqnarray}
        S^F(p) = i\, \left( \frac{\Lambda^+(\vec{p} \, )}
                                 {p_0-\omega+i\epsilon} \,
                 + \, \frac{\Lambda^-(\vec{p}\, )}
                                 {p_0+\omega-i\epsilon} \,
            \right) \gamma^0
\end{eqnarray}
which is suitable to perform the $q^0$
integration using the residue theorem. Then we substitute
$\vec{p}\rightarrow -\vec{p}$ in the second term of eq.(\ref{tmat}) and use
$\Gamma  (-\vec{p}\,)= -\gamma^0\Gamma  (\vec{p}\,)\gamma^0$  for
pseudoscalar mesons which follows from eq.(\ref{par}).
We use $\{\gamma^0,\varepsilon\!\!\! /_i\}_+=0$ for transversal photons
and define the quantities
\begin{eqnarray}
    \Theta^{\pm}(\vec{k}-\vec{p}\,)  :=
    \varepsilon\!\!\! /_2 \Lambda^{\pm}(\vec{k}-\vec{p}\,)
                             \gamma^0 \varepsilon\!\!\! /_1
  - \varepsilon\!\!\! /_1 \Lambda^{\pm}(\vec{k}-\vec{p}\,)
                             \gamma^0 \varepsilon\!\!\! /_2
\end{eqnarray}
with $\omega_k:=((\vec{k}-\vec{p}\,)^2+m^2)^{1/2}$.
Furthermore the product of the vertex function with the
projection operators $\Lambda^{\pm}$ is denoted as:
\begin{eqnarray}
 \Gamma^{+-}(\vec{p}\,)\, := \,\Lambda^{+}(\vec{p}\,)\gamma^0
\Gamma(\vec{p}\,)\gamma^0\Lambda^{-}(\vec{-p}\,)
\end{eqnarray}
and $\Gamma^{++}$ etc. analogously. In concrete calculations these
quantities are evaluated using a matrix formalism, which represents
the multiplication of $\Gamma$ with a Dirac matrix by a
transformation amongst the scalar amplitudes (see (\ref{repdm})).
After some calculation we obtain:
\begin{eqnarray}
T  &=&  -i \sqrt{3} (ie_q)^2
\, \int \!\!\! \frac{d^3p}{(2\pi)^3}
               \sum_{\pm}
\label{ttg} \\
&&\mbox{tr}\,\Bigg\{
       \mp\, \frac{ \Gamma ^{\mp\mp}\; \Theta^{\pm}
                                          (\vec{k}-\vec{p}\,)}
        {( M/2 \pm \omega_k\pm \omega )( - M/2\pm \omega_k \pm \omega )}
\nonumber \\ &&\;
       \mp\, \frac{ \Gamma ^{\mp\pm}\; \Theta^{\mp}
                                          (\vec{k}-\vec{p}\,)}
        {( M \pm 2\omega ) (M/2\pm \omega_k  \pm \omega )}
\nonumber \\ &&\;
       \mp\, \frac{ \Gamma ^{\mp\pm}\; \Theta^{\mp}
                                          (\vec{k}-\vec{p}\,)}
        {( M \pm 2\omega ) (M/2\pm \omega_k  \pm \omega )} \Bigg\}
\nonumber
\end{eqnarray}
where the summation $\pm$ runs over the upper
and lower sign.For $\Theta$ we obtain
\begin{eqnarray}
       \Theta^{\pm}(\vec{k}-\vec{p}\,)& = &
       i ( \vec{\varepsilon}_1\times\vec{\varepsilon}_2)
                 \cdot
\vec{\Sigma}^{\pm}(\vec{p}-\vec{k}\,)\hspace{1cm}
\end{eqnarray}
with
\begin{eqnarray}
     \vec{\Sigma}^{\pm} (\vec{p}-\vec{k}\,)
       & := & \gamma^5 \,\left( \vec{\gamma} \pm
          \frac{1}{\omega_k} (-\gamma^0 (\vec{p}-\vec{k}\,)
             + m\,\gamma^0 \vec{\gamma})\right)
\end{eqnarray}
To avoid the zeros in the denominator for $M>2m$ in eq.(\ref{ttg}) we
use the BS equation in the
form (\ref{9}) to obtain finally (with
$\vec{\Sigma}^0=\vec{\Sigma}^+-\vec{\Sigma}^-$):
\begin{eqnarray}
T  &=&  \sqrt{3} \;e_q^2\;
         ( \vec{\varepsilon}_1\times\vec{\varepsilon}_2)\cdot
                                           \vec{S}(\vec{k}\,)
    \;\;\;\;\mbox{with}\\
\vec{S}(\vec{k}\,)  &:=&
         \int \! \frac{d^3p}{(2\pi)^3}
       \;\frac{ 1}  {M^2/4 -(\omega_k +  \omega )^2}
\\    & & \mbox{tr} \,\Bigg\{
         \Gamma^{++}(\vec{p}\,)\;\vec{\Sigma}^-(\vec{p}-\vec{k}\,)
        - \Gamma^{--}(\vec{p}\,)\;\vec{\Sigma}^+(\vec{p}-\vec{k}\,)
\nonumber \\ &&\;
        -\frac{M}{M+2\omega}\,
         \Gamma^{-+}(\vec{p}\,)\;\vec{\Sigma}^0(\vec{p}-\vec{k}\,)
\nonumber \\ &&\;
        +\frac{i}{2\pi}(\frac{M}{2}+\omega_k+\omega)\,
         \Phi(\vec{p}\,)\;\vec{\Sigma}^0(\vec{p}-\vec{k}\,) \Bigg\}
\nonumber
\end{eqnarray}
As for transverse photons we have
$\vec{\varepsilon}_1\perp\vec{k}\perp\vec{\varepsilon}_2$, for
$\vec{k}= k \, \vec{e}_z$ the only nonvanishing component of
$\vec{\varepsilon}_1\times\vec{\varepsilon}_2$ is in z-direction and we
only need to calculate the z-component of $\vec{S}$. The standard
formula for the decay rate of a particle with mass $M$ then yields:
\begin{eqnarray}
      \Gamma(\pi^0,\eta\rightarrow 2\gamma) & = &
             \frac{3}{16\pi}\frac{\alpha^2\, \tilde{e}_q^4}{M}
                                        |S_z(k\vec{e}_z) |^2
\end{eqnarray}

\section{Conclusion} \label{V}
We have investigated the structure of the instantaneous BS-equation
(Salpeter equation) for the general case of unequal quark masses.
Furthermore we have developed a numerical scheme to solve
the Salpeter equation which enables the calculation of
mass spectra and Salpeter amplitudes. In order to test various models
beyond the mere reproduction of the mass spectra, we have further given
explicit formulas for the computation of weak meson decay constants
(\(f_{\pi},\,f_K\) etc.), the decay widths into two photons
and into an electron-positron pair.

Because of the relativistic kinematics,
the correct relativistic normalization of the amplitudes and the
dynamical treatment of the lower component \(\Phi^{--}\)
we expect the Salpeter equation to provide a framework
for quark models that is superior to other treatments like
the reduced Salpeter equation or the nonrelativistic quark model.
We will investigate an explicit quark model based on the Salpeter equation
in a subsequent paper \cite{Mue}.

\begin{appendix}
\section{Special Lorentz transformations} \label{A1}
Let \(\Lambda\) be a special Lorentz transformation
and \(g\) be the corresponding element of the covering group
\(SL(2,C)\), given by
\begin{equation}
g\,\sigma(x)\,g^{\dagger} = \sigma(\Lambda x)
\end{equation}
with \( \sigma(x)  =  x^{\mu} \sigma_{\mu}\) and
\( (\sigma_{\mu}) = (1,\vec{\sigma})\).
The transformation matrix \(S_g\) for Dirac spinors in the Weyl representation
is then given by
\begin{equation}
S_g =
\left( \begin{array}{*{2}{c}}
g & 0 \\
0 & (g^{\dagger})^{-1}
\end{array} \right)
\end{equation}
With the transformation matrix
\begin{equation}
B = \frac{1}{\sqrt{2}}\,
\left( \begin{array}{*{2}{c}}
1 & 1 \\
1 & -1
\end{array} \right)
\end{equation}
we can use \(S_g^D = B\,S_g^W\,B^{-1}\) to transform \(S_g\)
into the standard Dirac basis employed in this work.
The result reads
\begin{equation}
S_g =\frac{1}{2}\,
\left( \begin{array}{*{2}{c}}
g+(g^{\dagger})^{-1} & g-(g^{\dagger})^{-1} \\
g-(g^{\dagger})^{-1} & g+(g^{\dagger})^{-1}
\end{array} \right)
\end{equation}
Note that the relation
\(\Lambda^{\mu}_{\;\nu}\,\gamma^{\nu} = S_g^{-1}\,\gamma^{\mu}\,S_g\)
is automatically fulfilled. For a boost \(\Lambda\) with
\(\Lambda\,(M,\vec{0})=P\) and \(P^2=M^2\) one has explicitely
\begin{equation}
g = \left[\sigma \left(\frac{P}{M}\right) \right]^{\frac{1}{2}}
= \sqrt{\frac{M}{2(M+P^0)}}\,
\left(\,1+\sigma\left(\frac{P}{M}\right)\,\right)
\end{equation}
and for a 3-dimensional rotation \(\Lambda=R\) one has
\(g=u \in SU(2)\) and therefore
\begin{equation}
S_u =
\left( \begin{array}{*{2}{c}}
u & 0 \\
0 & u
\end{array} \right)
\end{equation}
in the Weyl basis as well as in the standard Dirac basis.

The transformation of a Dirac field operator \(\Psi(x)\)
is given by
\begin{eqnarray}
U_g\,\Psi(x)\,U_g^{-1} &=& S_g^{-1}\,\Psi(\Lambda x) \\
U_g\,\bar{\Psi}(x)\,U_g^{-1} &=& \bar{\Psi}(\Lambda x)\,S_g
\end{eqnarray}
with \(U_g\) being a unitary operator. Consider a state
with momentum \(P\) and \(P^2=M^2\), total angular momentum \(J\)
and 3-component \(M_J\) normalized as
\( \langle P',J', M_J' |\,P,J,M_J \rangle
= (2\pi)^3\,2P^0\,\delta^3(\vec{P}\,'-\vec{P})\,
\delta_{J'J}\,\delta_{M_J'M_J} \).
Then the action of \(U_g\)
on this state is given by
\begin{equation}
U_g\,|P,J,M_J\rangle = \sum_{M_J'}\,
|\Lambda P,J,M_J'\rangle\,D^J_{M_J'\,M_J}(u)
\end{equation}
where \(D^J_{M_J'\,M_J}(u)\) is a Wigner D-function
with \(D^{1/2}_{M_J'\,M_J}(u) = u_{M_J'\,M_J} \)
and \(u=[\sigma(\Lambda P/M)]^{1/2}\,g\,[\sigma(P/M)]^{1/2}\)
is the Wigner rotation.

{}From the definition of the BS-amplitude in Sec.\ref{IIA}
one can see by inserting the unity operator \(1=U_g^{-1}\,U_g\)
that the BS-amplitude transforms as
\begin{equation}
\chi^{JM_J}_P(p) =
\sum_{M_J'}\,S_g^{-1}\,\chi^{JM_J'}_{\Lambda P}(\Lambda p)
\,S_g\;D^J_{M_J'\,M_J}(u) \label{100}
\end{equation}
The BS-equation is compatible with this
transformation law for covariant kernels.

\section{Angular decomposition of the 2\(\times\)2-amplitudes}
Let \(\Lambda=R\) be a 3-dimensional rotation and \(g=u\)
be the corresponding matrix \(\in SU(2)\).
With eq.(\ref{100}) and the block matrix structure
of \(\Phi\) given in eq.(\ref{14}) as well as the relation
\((-i\sigma_2)\,u^{-1}\,i\sigma_2 = {}^t u\)
we find
\begin{eqnarray}
\lefteqn{ [\Phi^{++}(\vec{p})]_{JM_J}\,i\sigma_2} \label{110} \\
 &=&
\sum_{M_J'}\, [u \times u]\;
\{[\Phi^{++}(R^{-1}\vec{p})]_{JM_J'}\,i\sigma_2\}\;D^J_{M_J'M_J}(u^{-1})
\nonumber
\end{eqnarray}
and the same for the other amplitudes where we use the tensor notation
\( \{[u \times u]\,(\Phi^{++}\,i\sigma_2)\}_{m_1m_2} =
\sum_{m_1'm_2'}\,u_{m_1m_1'}\,u_{m_2m_2'}\,
(\Phi^{++}\,i\sigma_2)_{m_1'm_2'} \).
Therefore one can identify the indices \(m_1,\,m_2\)
of the matrix elements
\( \left[ \Phi^{++}\,i\sigma_2 \right]_{m_1m_2} \)
with the 3-components of the spin of the
quark and the anti\-quark. Let
\( (\chi_{SM_S})_{m_1m_2} = \langle 1/2\,m_1\,1/2\,m_2| SM_S \rangle \)
be the spin matrix of the two quarks coupled to the total spin \(S\).
Define
\(\varphi_{S\,q}\,i\sigma_2 := \chi_{S\,q}\), i.e.
\begin{equation}
\varphi_{0\,0} = \frac{1}{\sqrt{2}}\,1 \;\;\;,\;\;\;
\varphi_{1\,q} = \frac{1}{\sqrt{2}}\,\sigma_q
\end{equation}
or explicitely
\begin{eqnarray}
\varphi_{0\,0} = \frac{1}{\sqrt{2}} \,
\left( \begin{array}{*{2}{c}}
1&0\\0&1 \end{array} \right)
&\;\; ;\;\;&
\varphi_{1\,1} =
\left( \begin{array}{*{2}{c}}
0&-1\\0&0 \end{array} \right)
\\
\varphi_{1\,0} = \frac{1}{\sqrt{2}} \,
\left( \begin{array}{*{2}{c}}
1&0\\0&-1 \end{array} \right)
&\;\; ;\;\;&
\varphi_{1\,-1}=
\left( \begin{array}{*{2}{c}}
0&0\\1&0 \end{array} \right)
\end{eqnarray}
Then eq.(\ref{110}) implies that we can decompose
\(\Phi^{++},\,\Phi^{--}\) as
\begin{eqnarray}
\Phi^{++}(\vec{p}) &=&
 \sum_{L\,S}\,{\cal R}_{LS}^{(+)}(p) \,
 \left[Y_L(\Omega_p) \otimes \varphi_S \right]^J \nonumber \\
\Phi^{--}(\vec{p}) &=&
 \sum_{L\,S}\,{\cal R}_{LS}^{(-)}(p) \,
 \left[Y_L(\Omega_p) \otimes \varphi_S \right]^J \label{120}
\end{eqnarray}
with the spin \(S\) and the orbital angular momentum \(L\) coupled
to \(J\). We assume that the BS-kernel allows choosing
\({\cal R}_{LS}^{(+)}(p)\) to be a real function.
The sum goes over all values \(L,S\)
that are compatible with parity and charge parity of the bound state,
see below.

\section{Parity transformation of the BS-amplitude}
The Dirac field operator \(\Psi(x)\) and the bound state
with parity number \(\pi_P=\pm1\) transform under parity
transformation \(\hat{P}\) as
\begin{eqnarray}
U_P\,\Psi(x)\,U_P^{-1} &=& \gamma^0\,\Psi(\hat{P} x) \\
U_P\,\bar{\Psi}(x)\,U_P^{-1} &=& \bar{\Psi}(\hat{P} x)\,\gamma^0 \\
U_P\,|K,J,M_J,\pi_P \rangle &=& \pi_P\,|K,J,M_J,\pi_P \rangle
\end{eqnarray}
where \(K\) is the momentum of the bound state and
\(\hat{P}\,x = (x^0,-\vec{x})\). For the BS-amplitude this implies
\begin{equation}
\chi_K(p) = \pi_P\,\gamma^0\,\chi_{\hat{P}K}(\hat{P}p)\,\gamma^0
\end{equation}
For the block matrix structure one finds
\begin{equation}
\left( \begin{array}{*{2}{c}}
\Phi^{+-}(\vec{p}) & \Phi^{++}(\vec{p}) \\
\Phi^{--}(\vec{p}) & \Phi^{-+}(\vec{p})
\end{array} \right)
 = \pi_P\,
\left( \begin{array}{*{2}{c}}
\Phi^{+-}(-\vec{p}) & -\Phi^{++}(-\vec{p}) \\
-\Phi^{--}(-\vec{p}) & \Phi^{-+}(-\vec{p})
\end{array} \right)
\label{par}
\end{equation}
To be compatible with the angular decomposition eq.(\ref{120})
we find the well-known condition \(\pi_P=(-1)^{L+1}\).

\section{Charge conjugation}
Let \(|f_1\bar{f}_2,\,P \rangle\) be a \(q\bar{q}\)-bound state
with flavors \(f_1\) and \(f_2\) and momentum \(P\). The charge conjugation
then acts like
\begin{eqnarray}
U_C\,\Psi_{\alpha}(x)\,U_C^{-1} &=&
\sum_{\beta}\,(S_C)_{\alpha\beta}\;{}^t\Psi_{\beta}^+(x) \\
U_C\;{}^t\Psi^+_{\alpha}(x)\,U_C^{-1} &=&
\sum_{\beta}\,(S_C)_{\alpha\beta}\,\Psi_{\beta}(x) \\
U_C\,|f_1\bar{f}_2,\,P \rangle &=& |f_2\bar{f}_1,\,P \rangle
\end{eqnarray}
and we find for the BS-amplitude
\begin{equation}
\chi^P_{f_1f_2}(p) = -(S_C\gamma_0)\;{}^t\chi^P_{f_2f_1}(-p)\,(S_C\gamma^0)
\label{150} \end{equation}
with the matrix \(S_C\gamma^0\) given in the standard basis as
\begin{equation}
S_C\gamma^0 =
\left( \begin{array}{*{2}{c}}
0 & -i\sigma_2 \\
-i\sigma_2 & 0
\end{array} \right) \label{sc}
\end{equation}
Making the choice \(\eta_1=\eta_2=1/2\) implies
\begin{equation}
\left( \begin{array}{*{2}{c}}
\Phi_{f_1f_2}^{+-}(\vec{p}) &\Phi_{f_1f_2}^{++}(\vec{p}) \\
\Phi_{f_1f_2}^{--}(\vec{p}) &\Phi_{f_1f_2}^{-+}(\vec{p})
\end{array} \right) =
\left( \begin{array}{*{2}{c}}
\tilde{\Phi}_{f_2f_1}^{-+}(-\vec{p}) &
\tilde{\Phi}_{f_2f_1}^{++}(-\vec{p}) \\
\tilde{\Phi}_{f_2f_1}^{--}(-\vec{p}) &
\tilde{\Phi}_{f_2f_1}^{+-}(-\vec{p})
\end{array} \right)
\end{equation}
with \(\tilde{\Phi}^{++}=-i\sigma_2\,(^t\Phi^{++})\,i\sigma_2\).
With the angular decomposition eq.(\ref{120}) and
\( \tilde{\varphi}_{S\,q} = (-1)^S\,\varphi_{S\,q} \) one finds
\begin{equation}
\left[ {\cal R}_{LS}^{(\pm)}(p)\right]_{f_1f_2}
= (-1)^{L+S}\,
\left[ {\cal R}_{LS}^{(\pm)}(p)\right]_{f_2f_1}
\end{equation}
For an eigenstate \(|P,\,\pi_C\rangle\) of the charge conjugation
we have \(f_1=f_2\) and
\(U_C\,|P,\,\pi_C \rangle = \pi_C\,|P,\,\pi_C\rangle\)
which implies the well-known condition \(\pi_C=(-1)^{L+S}\).

To investigate the compatibility of eq.(\ref{150}) with the
BS-equation we use the relation
\( S_{f_i}^F(p_i) = -(S_C\gamma^0)\;{}^tS_{f_i}^F(-p_i)\,(S_C\gamma^0) \)
for the fermion propagator. Let \(\chi_{f_1f_2}(p)\) be a solution
of the BS-equation with \(f_1\) being the quark flavor and \(f_2\) being the
antiquark flavor. It is straightforward to show that \(\chi_{f_2f_1}(p)\)
as given in eq.(\ref{150}) is a solution of the BS-equation with interchanged
flavors.
For the Salpeter equation this implies that the solution
\(\psi_{f_1f_2}\) and the solution of the equation with
flavors interchanged \(\psi_{f_2f_1}\) are connected by
\begin{equation}
\psi_{f_1f_2}(\vec{p}) = -S_C\;{}^t \psi_{f_2f_1}(-\vec{p})\,S_C
\end{equation}

\section{Angular Decomposition of the 4$\times$4 Amplitude}
{}From the angular decomposition of the 2$\times$2 blocks and the
property under parity transformation we also obtain the structure of
the 4$\times$4 amplitudes.

With the spherical components $\sigma_m$ of the Pauli matrices we
define the two tensors
\begin{eqnarray}
     \gamma^{[0]}  =  \left( \begin{array}{cc} 1 & 0 \\
                                         0 & 1 \end{array} \right)
  \hspace{1cm}
    \gamma^{[1]}_{m_s}  =  \left( \begin{array}{cc} 0 & \sigma_{m_s} \\
                                   -\sigma_{m_s} & 0 \end{array} \right),
\end{eqnarray}
which are again Dirac matrices. We obtain the following representation
for the BS amplitudes in the rest frame
\begin{eqnarray}
\lefteqn{
     \chi^{JM_J,\pi_P}(p)
           = \sum_{i=1}^8 \;\;\chi_i (p^0,|\vec{p}\,|)\,
                         \; \Gamma_i(\Omega_p) =
} \label{adchi} \\
      & := & \;\sum^8_{i=1}
     \;\chi _{l_i,s_i}^{a_i,g_i}(p^0,|\vec{p}\,|)
     \;(\gamma^5)^{a_i}  \;(\gamma^0)^{g_i}
       \left[ Y_{l_i}(\Omega_p)\times \gamma^{[s_i]}\right]^J_{M_J}
\nonumber
\end{eqnarray}
The sum over i covers all the
values of $s_i,a_i,g_i\in \{ 0,1 \}$ and
$l_i\in\{J,J\pm 1\}$ that are compatible with
\((-1)^{s_i+l_i+a_i}=\pi_P \), where \(\pi_P\) is the parity of the
meson.
In general we have eight scalar amplitudes
$\chi_i(p^0,|\vec{p}\,|)$ (except for mesons with spin 0, where there
are only four).

The same structure holds for the Salpeter amplitudes $\Phi$, the
vertex functions $\Gamma$ and any amplitude that is built out of these by
multiplication with e.g. $\gamma^0, \vec{\gamma}\Omega_p$ or
$\Lambda^{\pm}(\vec{p}\,)$. It is therefore convenient to introduce a
matrix formalism for such multiplications which is useful for the
calculation of decays or transitions and simplifies the Dirac algebra.

{}From eq.(\ref{adchi}) we may write the amplitude in the form
\begin{eqnarray}
       \Phi(\vec{p}\,) = \sum_{i=1}^8\;\; \Phi_i (|\vec{p}\,|)\,
                         \; \Gamma_i(\Omega_p)
\end{eqnarray}
so that we can express the multiplication from the left or right hand
side with a matrix $\tilde{\gamma}\in\{\gamma^0,
\vec{\gamma}\vec{p},...\}$ by a transformation $D_L(\tilde{\gamma})$ or
$D_R(\tilde{\gamma})$ in the 8 dimensional space of the radial amplitudes:
\begin{eqnarray}
   \tilde{\gamma}\;\Phi (\vec{p}\,)=\sum_{i,k=1}^8\;  \Phi_i (|\vec{p}\,|)\,
                    \;    D_L(\tilde{\gamma}\,)_{ik}\; \Gamma_k(\Omega_p)
\label{repdm}
\end{eqnarray}
and analogously with $D_R$ for multiplication from the right side. We also
use the following formula to calculate the trace in Dirac space
and the angular integration:
\begin{eqnarray}
     \int\!\! d\Omega_p\; tr\; \left\{\;\Gamma^{\dagger}_i (\Omega_p)
               \; \Phi(\vec{p}\,)\;\right\}\; = \;4\;\Phi_i(|\vec{p}\,|)
\end{eqnarray}

\end{appendix}

\vspace{1cm}

\begin{figure}[ht]
  \centering
  \leavevmode
  \epsfxsize=0.50\textwidth
  \epsffile{fig2.eps}
\vspace{0.5cm}
\caption{The normalization condition}
\label{vnorm}
\end{figure}

\begin{figure}[ht]
  \centering
  \leavevmode
  \epsfxsize=0.33 \textwidth
  \epsffile{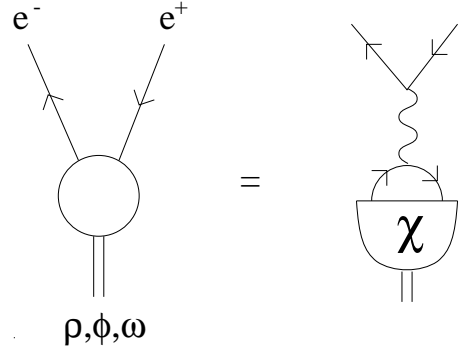}
\vspace{0.5cm}
\caption{Leptonic decays in the Mandelstam formalism up to lowest
order in the electromagnetic coupling constant}
\label{mandel}
\end{figure}

\begin{figure}[ht]
  \centering
  \leavevmode
  \epsfxsize=0.38\textwidth
  \epsffile{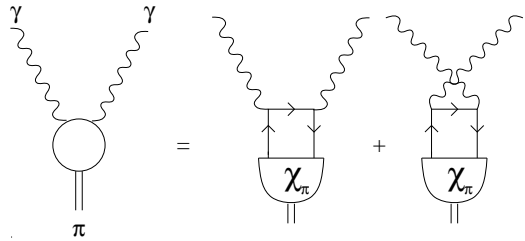}
\vspace{0.5cm}
\caption{The decay $\pi^0,\eta\rightarrow 2\gamma$}
\label{twogamma}
\end{figure}
\noindent

\end{document}